\documentclass{article}

\usepackage{microtype}
\usepackage{graphicx}
\usepackage{booktabs}
\usepackage{xcolor}
\usepackage{bm}
\usepackage{amsmath}
\usepackage{amsfonts}
\usepackage{dsfont}
\usepackage{subcaption}
\usepackage{tikz, enumerate}
\usetikzlibrary{arrows,intersections,positioning,shapes.arrows,decorations.pathreplacing}
\usepackage{pgfplots}
\usepackage{hyperref}

\DeclareMathOperator*{\argmax}{arg\!max}
\DeclareMathOperator*{\argmin}{arg\!min}

\newcommand\q{{\bm{q}}}
\newcommand\s{q}
\newcommand\Q{\mathcal{Q}}
\newcommand\ag{\bm{\alpha}}
\newcommand{\bth}{\boldsymbol{\theta}} 
 
\newcommand{\bx}{\boldsymbol{x}} 
\newcommand{\tx}{\textbf{x}}

\newcommand{\ty}{\textbf{y}}

\newcommand\bleu[1]{\textcolor{black}{#1}}

\def\mybar#1#2#3{%%
\begin{tabular}{@{}l@{}} {\scriptsize #1} \vspace*{-0.2cm} \\ {\scriptsize #2} \vspace*{-0.2cm} \\ {\scriptsize #3}\end{tabular} & \resizebox{.05\textwidth}{0.5cm}{\begin{tabular}{@{}l@{}}{\color{green}\rule[-6pt]{#1bp}{10pt}} \\ {\color{orange}\rule[0pt]{#2bp}{10pt}} \vspace*{-0.1cm} \\ {\color{red}\rule[10pt]{#3bp}{10pt}} \end{tabular}}}

\def\myobar#1#2#3{%%
\begin{tabular}{@{}l@{}} {\scriptsize #1} \vspace*{-0.2cm} \\ {\scriptsize #2} \vspace*{-0.2cm} \\ {\scriptsize #3}\end{tabular} & \resizebox{.05\textwidth}{0.5cm}{\begin{tabular}{@{}l@{}}{\color{orange}\rule[-6pt]{#1bp}{10pt}} \\ {\color{green}\rule[0pt]{#2bp}{10pt}} \vspace*{-0.1cm} \\ {\color{orange}\rule[10pt]{#3bp}{10pt}}\end{tabular}}}

\usepackage[accepted]{modded_icml2019}

\icmltitlerunning{Feature quantization for parsimonious and interpretable predictive models}

\pgfplotsset{compat=1.14}

\begin{document}

\twocolumn[
\icmltitle{Feature quantization for parsimonious and interpretable predictive models}

\begin{icmlauthorlist}
\icmlauthor{Adrien Ehrhardt}{cacf,inria}
\icmlauthor{Christophe Biernacki}{inria}
\icmlauthor{Vincent Vandewalle}{inria,soins}
\icmlauthor{Philippe Heinrich}{painleve}
\end{icmlauthorlist}

\icmlaffiliation{cacf}{Cr\'edit Agricole Consumer Finance, Roubaix, France}
\icmlaffiliation{inria}{Inria, Villeneuve d'Ascq, France}
\icmlaffiliation{soins}{EA2694 Sant\'e publique: \'epid\'emiologie et qualit\'e des soins, Univ. Lille, Lille, France}
\icmlaffiliation{painleve}{UMR 8524
Laboratoire Paul Painlev\'e, Univ. Lille, Lille, France}

\icmlcorrespondingauthor{Adrien Ehrhardt}{adrien.ehrhardt@inria.fr}
\icmlkeywords{discretization, quantization, factor, grouping, credit scoring, Machine Learning, ICML}

\vskip 0.3in
]

\printAffiliationsAndNotice{}

\begin{abstract}
For regulatory and interpretability reasons, logistic regression is still widely used. To improve prediction accuracy and interpretability, a preprocessing step quantizing both continuous and categorical data is usually performed: continuous features are discretized and, if numerous, levels of categorical features are grouped. An even better predictive accuracy can be reached by embedding this quantization estimation step directly into the predictive estimation step itself. But doing so, the predictive loss has to be optimized on a huge set. To overcome this difficulty, we introduce a specific two-step optimization strategy: first, the optimization problem is relaxed by approximating discontinuous quantization functions by smooth functions; second, the resulting relaxed optimization problem is solved \textit{via} a particular neural network. The good performances of this approach, which we call \textit{glmdisc}, are illustrated on simulated and real data from the UCI library and Cr\'edit Agricole Consumer Finance (a major European historic player in the consumer credit market).
\end{abstract}

\section{Motivation} \label{sec:motivation}

As stated by~Hosmer et al.~\yrcite{hosmer2013applied}, in many application contexts (credit scoring, biostatistics, {\it etc.}), logistic regression is widely used for its simplicity, decent performance and interpretability in predicting a binary outcome given predictors of different types (categorical, continuous). However, to achieve  higher interpretability, continuous predictors are sometimes discretized so as to produce a ``scorecard'', \textit{i.e.}\ a table assigning a grade to an applicant in credit scoring (or a patient in biostatistics, {\it etc.}) depending on its predictors being in a given interval.
Discretization is also an opportunity for reducing the (possibly large) modeling bias which can appear in logistic regression as a result of the linearity assumption on the continuous predictors in the model. Indeed, this restriction can be overcome by approximating the true predictive mapping with a step function where the tuning of the steps and of their sizes allows more flexibility. However, the resulting increase of the number of parameters can lead to an increase of their variance (overfitting) as shown by Yang \& Webb~\yrcite{yang2009discretization}. Thus, a precise tuning of the discretization procedure is required. Likewise when dealing with categorical features which take numerous levels, their respective regression coefficients suffer from this high variance phenomenon. A straightforward solution formalized by Maj-Ka\'nska et al.~\yrcite{maj2015delete} is to merge their factor levels which leads to less coefficients and therefore less variance. 

From now on, the generic term quantization will stand for both discretization of continuous features and level grouping of categorical ones. Its aim is to improve the prediction accuracy but it suffers from yielding a highly combinatorial optimization problem whatever the predictive criterion used to select the best quantization. The present work proposes a strategy to overcome these combinatorial issues by invoking a relaxed alternative of the initial quantization problem leading to a simpler estimation problem since it can be easily optimized by a specific neural network. This relaxed version serves as a plausible quantization provider related to the initial criterion after a classical thresholding (\textit{maximum a posteriori}) procedure.

The outline of this work is the following. In the next section, we formalize both continuous and categorical quantization. Selecting the best quantization in a predictive setting is reformulated as a model selection problem on a huge discrete space. In Section~\ref{sec:proposal}, a particular neural network architecture is used to optimize a relaxed version of this criterion and propose good quantization candidates. Section~\ref{sec:experiments} is dedicated to numerical experiments on both simulated and real data from the field of Credit Scoring, highlightening the good results offered by the use of this new method without any human intervention. A final section concludes the work by stating also new challenges.

\section{Quantization as a combinatorial challenge} \label{sec:model_selection}

\subsection{Quantization: definition}

The quantization procedure consists in turning a $d$-dimensional raw vector of continuous and/or categorical features $\bm{x} = (x_1, \ldots, x_d)$ into a $d$-dimensional categorical vector via a component wise mapping $\q=(\q_j)_1^d$:
\[\q(\bm{x})=(\q_1(x_1),\ldots,\q_d(x_d)),\]
where each of the $\q_j$'s is a vector of $m_j$ dummies: 
\begin{equation}\label{eq:qj}
\s_{j,h}(\cdot) =  1 \text{ if } x_j \in C_{j,h}, 0 \text{ otherwise, } 1 \leq h \leq m_j,
\end{equation}
where $m_j$ is an integer and the sets $C_{j,h}$ are defined with respect to each feature type as we describe just below.

\subsubsection{Raw continuous features case} If $x_j$ is a continuous component of $\bm{x}$, quantization $\q_j$ has to perform a discretization of $x_j$ and the $C_{j,h}$s, $1\le h\le m_j$, are contiguous intervals  
\begin{equation}\label{eq:Cjhcont}
C_{j,h}=(c_{j,h-1},c_{j,h}]
\end{equation}
where $c_{j,1},\ldots,c_{j,m_j-1}$ are increasing numbers called cutpoints, $c_{j,0}=-\infty$ and $c_{j,m_j}=+\infty$.

For example, the quantization of the unit segment in thirds would be defined as $m_j=3$, $c_{j,1} = 1/3$, $c_{j,2} = 2/3$ and subsequently $\q_j(0.1) = (1,0,0)$.

\subsubsection{Raw categorical features case} 

If $x_j$ is a categorical component of $\bm{x}$, quantization $\q_j$ consists in grouping levels of $x_j$ and the $C_{j,h}$s form a partition of the set, say $\{1,\ldots,l_j\}$, of levels of $x_j$: 
%\begin{equation}\label{eq:Cjhcat}
\begin{equation*}%\label{eq:Cjhcat}
\bigsqcup_{h=1}^{m_j}C_{j,h}=\{1,\ldots,l_j\}.
%\end{equation}
\end{equation*}
For example, the grouping of levels encoded as ``1'' and ``2'' would yield $C_{j,1} = \{1,2\}$ such that $\q_j(1) = \q_j(2) = (1,0,\ldots,0)$.

\subsubsection{Notations for the quantization family}

\bleu{In both continuous and categorical cases, keep in mind that $m_j$ is the dimension of $\q_j$. For notational convenience, the (global) order of the quantization $\q$ is set as 
\[|\q|=\sum_{j=1}^d m_j.\]
The space where quantizations $\q$ live (resp. $\q_j$) will be denoted by $\Q_{\bm{m}}$ in the sequel (resp. $\Q_{j,m_j}$), when the number of levels $\bm{m} = (m_j)_1^d$ is fixed. Since it is not known, the full model space is $\Q = \cup_{m \in \mathbb{N}_\star^{d}} \Q_{\bm{m}}$.}

\subsubsection{Literature review}

The current practice of quantization is prior to any predictive task, thus ignoring its consequences on the final predictive ability. It consists in optimizing a heuristic criterion, often either totally unrelated (unsupervised methods) or partially related (supervised methods) to the predictive task, and mostly univariate (each feature is quantized irrespective of other features' values). The cardinality of the quantization space $\Q$ can be calculated explicitely w.r.t.\ $d$, $(m_j)_1^d$ and, for categorical features, $l_j$. It is huge (see a more precise illustration of this combinatorial challenge in Section~\ref{subsubsec:model_selection}), so that a greedy approach is intractable and such heuristics are needed.
Many algorithms have thus been designed and a review of approximatively 200 discretization strategies, gathering both criteria and related algorithms, can be found in \cite{ramirez2016data}. For factor levels grouping, we found no such taxonomy, but some discretization methods, \textit{e.g.}\ $\chi^2$ independence test-based methods can be naturally extended to this type of quantization, which is for example what the CHAID algorithm, proposed by Kass~\yrcite{kass1980exploratory} and applied to each categorical feature, relies on.

\subsection{Quantization embedded in a predictive process}

\subsubsection{Logistic regression on quantized data}

Quantization is a widespread preprocessing step to perform a learning task consisting in predicting, say, a binary variable $y\in\{0,1\}$, from a quantized predictor  $\q(\bm{x})$, through, say, a parametric conditional distribution $p_{\bth}(y|\q(\bm{x}))$ like logistic regression. Considering quantized data instead of raw data has a double benefit. First, the quantization order $|\q|$ acts as a tuning parameter for controlling the model's flexibility and thus the bias/variance trade-off of the estimate of the parameter $\bth$ (or of its predictive accuracy) for a given dataset. This claim becomes clearer with the example of logistic regression we focus on, as a still very popular model for many practitioners. It is classically described by
\begin{equation}
    \label{eq:reglogq}
\ln \left( \dfrac{p_{\bth}(1|\q(\bm{x}))}{1 - p_{\bth}(1|\q(\bm{x}))} \right) = \theta_0 + \sum_{j=1}^d \bth_j' \cdot \q_j(x_j),
\end{equation}
where $\bth = (\theta_{0},(\bth_j)_1^d) \in \mathbb{R}^{|\q|+1}$ and $\bth_j = (\theta_{j}^{1},\dots,\theta_{j}^{m_j})$ with $\theta_{j}^{m_j} = 0$, $j=1 \ldots d$, for identifiability reasons.
Second, at the practitioner level, the previous tuning of $|\q|$ through each feature's quantization order $m_j$, especially when it is quite low, allows an easier interpretation of the most important predictor values involved in the predictive process. Denoting the dataset by $(\tx,\ty)$, with $\tx=(\bm{x}_1,\ldots,\bm{x}_n)$, $\ty=(y_1,\ldots,y_n)$ and $n$ the sample size, the log-likelihood 
\begin{equation}
\label{eq:lq}
\ell_{\q}(\bth ; (\tx,\ty))=\sum_{i=1}^n \ln p_{\bth}(y_i|\q(\bm{x}_i))
\end{equation}
provides a maximum likelihood estimator $\hat{\bth}_\q$ of $\bth$ for a given quantization $\q$. For the rest of the paper, the approach is exemplified with logistic regression as $p_{\bth}$ but it can be applied to any other predictive model, as will be recalled in the concluding Section~(\ref{sec:conclusion}).

\subsubsection{Quantization as a model selection problem} \label{subsubsec:model_selection}

As dicussed in the previous section, and emphasized in the literature review, quantization is often a preprocessing step; however, quantization can be embedded directly in the predictive model. Continuing our logistic example, a standard information criterion such as the BIC~\cite{BIC} can be used to select the best quantization $\q$:
\begin{align}
    \label{eq:BICq}
    \hat{\q} & = \argmin_{\q \in \Q} \mbox{BIC}(\hat{\bth}_\q) \\ \nonumber
    & = \argmin_{\q \in \Q}\left\{-2\ell_\q(\hat{\bth}_\q ; (\tx,\ty)) + \nu_\q \ln (n)\right\}
\end{align}
where $\nu_\q$ is the number of continuous parameters to be estimated in the $\bth$-parameter space. \bleu{We shall insist here on the fact that choosing the BIC as our model selection tool is unrelated to the proposed algorithm. The practitioner can swap this criterion with any other information criterion on training data such as AIC~\cite{akaike1973information} or, as \textit{Credit Scoring} people like, the Gini index on a test set.} Note however that, regardless of the criterion used, an exhaustive search of $\hat{\q}\in\Q$ is an intractable task due to its highly combinatorial nature. For example, with $d=10$ categorical features with $l_j = 4$ levels each, $|\Q|$ is given by the sum of the Stirling numbers of the second kind over $m_j=1 \dots l_j$ to the power $d$, which is approximately \ $6 \cdot 10^{11}$. Anyway, the optimization in~(\ref{eq:BICq}) requires a new specific strategy, which is the main contribution of the present work, and that we describe in the next section.

\subsubsection{Remark on model identifiability}

The shifting of cutpoints (\ref{eq:Cjhcont}) anywhere strictly between two successive raw values of a given continuous feature induce the same quantization. Thus, the identifiability of such quantizations is obtained from the dataset $\tx$ by fixing arbitrary cutpoints between successive data values, feature by feature.

\section{The proposed neural network-based quantization}
\label{sec:proposal}

\subsection{A relaxation of the optimization problem}

In this section, we propose to relax the constraints on $\q_j$ to simplify the search of $\hat{\q}$. Indeed, the derivatives of $\q_j$ are zero almost everywhere and consequently a gradient descent cannot be directly applied to find an optimal quantization.

\subsubsection{Smooth approximation of the quantization mapping}

A classical approach is to replace the binary functions $q_{j,h}$ (see Equation (\ref{eq:qj}))  by smooth parametric ones  with a simplex condition, namely with $\bm{\alpha}_j=(\bm{\alpha}_{j,1},\ldots, \bm{\alpha}_{j,m_j})$:
\begin{equation*}
    \q_{\ag_j}(\cdot)=\left(q_{\ag_{j,h}}(\cdot)\right)_{h=1}^{m_j} \text{ with } \begin{cases} \sum_{h=1}^{m_j}q_{\ag_{j,h}}(\cdot)=1, \\ 0 \leq q_{\ag_{j,h}}(\cdot) \leq 1, \end{cases}
\end{equation*}
where functions $q_{\ag_{j,h}}(\cdot)$, properly defined hereafter for both continuous and categorical features, represent a fuzzy quantization in that, here, each level $h$ is weighted by $\s_{\ag_{j,h}}(\cdot)$ instead of being selected once and for all as in (\ref{eq:qj}). The resulting fuzzy quantization for all components depends on the global parameter $\bm{\alpha} = (\bm{\alpha}_1, \ldots, \bm{\alpha}_d)$ and is denoted by $\q_{\ag}(\bx)=\left(\q_{\ag_j}(x_j)\right)_{j=1}^d \in \widetilde{Q}$.
%\PH{non, le cdot n'est pas le meme dans l'egalite}.
This approximation will be justified in Section~\ref{subsubsec:justif}.

{\bf For continuous features}, we set for $\bm{\alpha}_{j,h} = (\alpha^0_{j,h},\alpha^1_{j,h}) \in \mathbb{R}^2$
\[\s_{\ag_{j,h}}(\cdot) = \frac{\exp(\alpha^0_{j,h} + \alpha^1_{j,h}  \cdot)}{\sum_{g=1}^{m_j} \exp(\alpha^0_{j,g} + \alpha^1_{j,g}  \cdot)}.\]
%where $\ag_{j,m_j}$ is set to $(0,0)$ for identifiability reasons.

{\bf For categorical features}, we set for $\bm{\alpha}_{j,h}=\left(\alpha_{j,h}(1),\ldots, \alpha_{j,h}(l_j)\right) \in \mathbb{R}^{l_j}$
\[\s_{\ag_{j,h}}(\cdot) = \frac{\exp\left(\alpha_{j,h}(\cdot)\right)}{\sum_{g=1}^{m_j} \exp\left(\alpha_{j,g}(\cdot)\right)}.\]
%where $l_j$ is the number of levels of the categorical feature $x_j$.

\subsubsection{Parameter estimation}

With this new fuzzy quantization, the logistic regression for the predictive task is then expressed as
\begin{equation}
    \label{eq:reglogqa}
    \ln \left( \dfrac{p_{\bth}(1|\q_{\ag} (\bm{x}))}{1 - p_{\bth}(1|\q_{\ag} (\bm{x}))} \right) = \theta_0 + \sum_{j=1}^d {\bth_j' \cdot \q_{\ag_{j}}(x_j)},
\end{equation}
where $\q$ has been replaced by $\q_{\ag}$ from Equation~(\ref{eq:reglogq}).
Note that as $\q_{\ag}$ is a sound approximation of $\q$ (see Section~\ref{subsubsec:justif}), this logistic regression in $\q_{\ag}$ is consequently a good approximation of the logistic regression in $\q$ from Equation~(\ref{eq:reglogq}). The relevant log-likelihood is here 
\begin{equation}
    \label{eq:lqa}
    \ell_{\q_{\bm{\alpha}}}(\bth ; (\tx,\ty))=\sum_{i=1}^n \ln p_{\bth}(y_i|\q_{\bm{\alpha}}(\bm{x}_i))
\end{equation}
and can be used as a tractable substitute for (\ref{eq:lq}) to solve the original optimization problem (\ref{eq:BICq}), where now both $\ag$ and $\bth$ have to be estimated, which is discussed in the next section. We wish to maximize the log-likelihood (\ref{eq:reglogqa}) which would yield parameters $(\hat{\ag},\hat{\bth})$; 
%\bleu{these are consistent if the model is well-specified (\textit{i.e.}\ there is a ``true'' quantization under classical regularity conditions).}
To ``push'' $\widetilde{\Q}$ further into $\Q$, we deduce ${\q}^{\text{MAP}}$ from a \textit{maximum a posteriori} procedure applied to $\q_{\hat{\ag}}$:
\begin{equation}
    \label{eq:ht}
    \hat{\s}_{j,h}^{\text{MAP}}(x_j) = 1 \text{ if } h = \argmax_{1 \leq h' \leq m_j} \s_{\hat{\ag}_{j,h'}}, 0 \text{ otherwise.}
\end{equation}
If there are two levels $h$ that satisfy (\ref{eq:ht}), we simply take the level that corresponds to smaller values of $x_j$ to be in accordance with the definition of $C_{j,h}$ in Equation~(\ref{eq:Cjhcont}). This {\it maximum a posteriori} principle are exemplified in Figure~\ref{fig:MAP} on simulated data by the plain vertical lines (see Section~\ref{sec:experiments}).

\subsubsection{Validity of the relaxation} \label{subsubsec:justif}

From a deterministic point of view, we have $\Q \subset \widetilde{\Q}$: First, the \textit{maximum a posteriori} step~(\ref{eq:ht}) produces contiguous intervals (\textit{i.e.}\ there exists $C_{j,h}$; $1 \leq j \leq d$, $1 \leq h \leq m_j$, s.t.\ ${\q}^{\text{MAP}}$ can be written as in~\ref{eq:qj}) \cite{same2011model}. Second, in the continuous case, the higher $\alpha_{j,h}^1$, the less smooth the transition from one quantization $h$ to its ``neighbor''\footnotemark[1] $h+1$, whereas $\dfrac{\alpha_{j,h}^0}{\alpha_{j,h}^1}$ controls the point in $\mathbb{R}$ where the transition occurs \cite{chamroukhi2009regression}. Concerning the categorical case, the rationale is even simpler as $q_{\lambda \alpha_{j,h}}(x_j) \to 1 \text{ if } h = \argmax_{h'} q_{\alpha_{j,h'}}(x_j), 0 \text{ otherwise}$ as $\lambda \to +\infty$~\cite{reverdy2016parameter}.

From a statistical point of view, under standard regularity conditions and with a suitable estimation procedure (see later for the proposed estimation procedure), the maximum likelihood framework ensures the consistency of $(\q_{\hat{\ag}}, \hat{\bth})$ towards $(\q,\bth)$. This is further ensured by the \textit{maximum a posteriori} step~(\ref{eq:ht}).

However, and as is usual, the log-likelihood $\ell_{\q_{\ag}}(\bth,(\tx,\ty))$ cannot be directly maximized w.r.t.\ $(\ag,\bth)$, so that we need an iterative procedure. To this end, the next section introduces a neural network of particular architecture.

From an empirical point of view, we will see in Section~\ref{sec:experiments} and in particular in Figure~\ref{fig:MAP}, that the smooth approximation $\q_{\ag}$ converges towards ``hard'' quantizations\footnotemark[1] $\q$.

\footnotetext[1]{Up to a permutation on the labels $h=1 \ldots m_j$ to recover the ordering in $C_{j,h}$ (see Equation (\ref{eq:Cjhcont})).}

\subsection{A neural network-based estimation strategy} \label{sec:estim}

\subsubsection{Neural network architecture}

To estimate parameters $\ag$ and $\bth$ in model (\ref{eq:reglogqa}), a particular neural network architecture can be used. \bleu{We shall insist that this network is only a way to use common deep learning frameworks, namely Tensorflow~\cite{tensorflow2015-whitepaper} through the high-level API Keras~\cite{chollet2015keras} instead of building a gradient ascent algorithm from scratch to optimize~\eqref{eq:lqa}.} The most obvious part is the output layer that must produce $p_{\bth}(1|\q_{\ag}(\bm{x}))$ which is equivalent to a densely connected layer with a sigmoid activation $\sigma (\cdot)$.

For a continuous feature $x_j$ of $\bm{x}$, the combined use of $m_j$ neurons including affine transformations and softmax activation obviously yields $\q_{\ag_{j}}(x_j)$. Similarly, an input categorical feature $x_j$ with $l_j$ levels is equivalent to $l_j$ binary input neurons (presence or absence of the factor level). These $l_j$ neurons are densely connected to $m_j$ neurons without any bias term and a softmax activation. The softmax outputs are next aggregated via the summation in model (\ref{eq:reglogqa}), say $\Sigma_{\bth}$ for short, and then the sigmoid function $\sigma$ gives the final output. All in all, the proposed model is straightforward to optimize with a simple neural network, as shown in Figure~\ref{fig:nn}.

\def\layersep{2.5cm}

\begin{figure*}[ht!]
\centering
\begin{tikzpicture}[shorten >=1pt,->,draw=black!50, node distance=\layersep]
    \tikzstyle{every pin edge}=[<-,shorten <=1pt]
    \tikzstyle{neuron}=[circle,fill=black!25,minimum size=17pt,inner sep=0pt]
    \tikzstyle{input neuron}=[neuron, fill=green!50];
    \tikzstyle{output neuron}=[neuron, fill=red!50];
    \tikzstyle{hidden neuron}=[neuron, fill=blue!50];
    \tikzstyle{annot} = [text width=4em, text centered]
    \tikzstyle{annotrectangle} = [text width=8em, text centered]

        \node[input neuron, pin=left:continuous value $x_j$] (I-1) at (0,-1) {};
        
        \node[input neuron, pin=left:categorical value $1$] (I-2) at (0,-2) {};
        \node[input neuron, pin=left:$\vdots$] (I-3) at (0,-3) {};
        \node[input neuron, pin=left:categorical value $l_j$] (I-4) at (0,-4) {};

    % Draw the hidden layer nodes
    \foreach \name / \y in {1,...,2}
        \path[yshift=0.5cm]
            node[hidden neuron] (H-\name) at (\layersep,-\y cm) {soft};

    \foreach \name / \y in {3,...,4}
        \path[yshift=0.5cm]
            node[hidden neuron] (H-\name) at (\layersep,-\y cm) {soft};
            
    % Draw the sum layer node 
    
    \node[neuron, right of=H-2] (S) {$\Sigma_{\bth}$};

    % Draw the output layer node
    
    \node[output neuron,pin={[pin edge={->}]right:output}, right of=S] (O) {$\sigma$};
    
    %\node[output neuron,pin={[pin edge={->}]right:Output}, right of=H-2] (O) {$\sigma(\cdot)$};

    % Connect every node in the input layer with every node in the
    % hidden layer.
%    \foreach \source in {1,...,4}
        \foreach \dest in {1,2}
            \path (I-1) edge (H-\dest);

        \foreach \dest in {3,4}
            \path (I-2) edge (H-\dest);
        \foreach \dest in {3,4}
            \path (I-3) edge (H-\dest);
        \foreach \dest in {3,4}
            \path (I-4) edge (H-\dest);

        % \foreach \dest in {5,6}
        %     \path (I-3) edge (H-\dest);

    % Connect every node in the hidden layer with the output layer
    \foreach \source in {1,...,4}
        \path (H-\source) edge (S);
        
    % connect Sigma with sigma
    \path (S) edge (O);

    % Annotate the layers
    \node[annot,above of=H-1, node distance=1cm] (hl) {softmax layer};
    \node[annot,above of=I-1,node distance=1cm] {weights $\ag_j$};
    %\node[annot,right of=hl] (s) {};
    \node[annot, below of=O, node distance=1cm] (s) {sigmoid function};
    \node[annot, below of=S,node distance=1cm] {summation function};
    
    \draw [orange] (2,0) rectangle (3,-1.9);
    % \draw [red] (2,-2) rectangle (3,-4);
    
    \node[annotrectangle,right of=H-1, node distance=1.5cm] {soft outputs $\q_{\ag_j}(x_j)$}; 

\end{tikzpicture}
\caption{Proposed shallow architecture to maximize (\ref{eq:lqa}).}
\label{fig:nn}
\end{figure*}
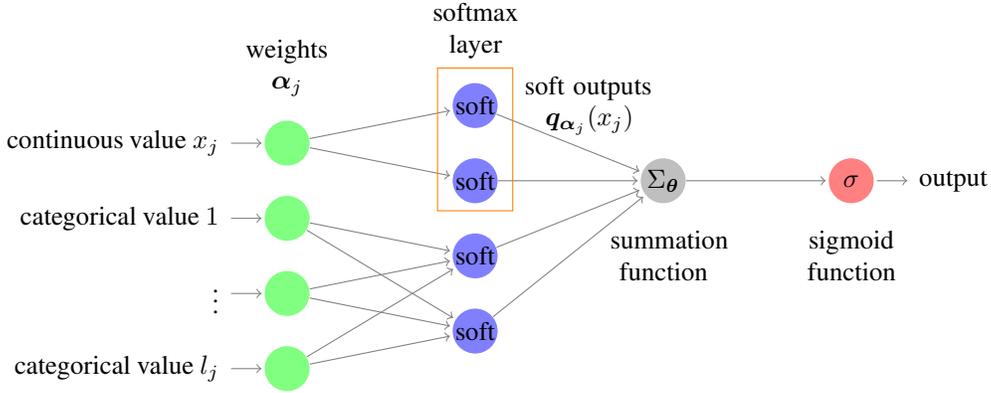

\subsubsection{Stochastic gradient descent as a quantization provider} \label{subsubsec:stochastic}

By relying on stochastic gradient ascent, the smoothed likelihood (\ref{eq:lqa}) can be maximized over $\left(\ag, \bth \right)$. Due to its convergence properties~\cite{bottou2010large}, the results should be close to the maximizers of the original likelihood (\ref{eq:lq}) if the model is well-specified, when there is a true underlying quantization. However, in the mis-specified model case, there is no such guarantee. Therefore, to be more conservative, we evaluate at each training epoch $(t)$ the quantization ${\q}^{\text{MAP}(t)}$ resulting from the \textit{maximum a posteriori} procedure explicited in Equation~(\ref{eq:ht}), then classically estimate the logistic regression parameter \textit{via} maximum likelihood, as done in Equation~(\ref{eq:lq}):
\[\hat{\bth}^{(t)} = \argmax_{\bth} \ell_{{\q}^{\text{MAP}(t)}}(\bth; (\tx,\ty))\]
and the resulting $\mbox{BIC}(\hat{\bth}^{(t)})$ as in (\ref{eq:BICq}). If $T$ is a given maximum number of iterations of the stochastic gradient ascent algorithm, the quantization retained at the end is then determined by the optimal epoch
\begin{equation} \label{eq:opt_epoch}
t_*=\argmin_{t\in \{1,\ldots, T\}} \mbox{BIC}(\hat{\bth}^{(t)}).
\end{equation}
\bleu{The number of iterations $T$ can be seen as a computational budget: contrary to classical early stopping rules (\textit{e.g.}\ based on validation loss) used in neural network fitting, this network only acts as a stochastic quantization provider for~\eqref{eq:opt_epoch} which will naturally prevent overfitting. We reiterate that, in~\eqref{eq:opt_epoch}, the BIC can be swapped for the user's favourite model choice criterion. Lots of optimization algorithms for neural networks have been proposed, which all come with their hyperparameters. We chose the ``RMSProp'' method, which showed good results, is one of the standard methods, and tuned only its learning rate.}

\subsubsection{Choosing an appropriate number of levels} \label{subsubsec:choosing}

The number of intervals or factor levels $\boldsymbol{m} = (m_j)_1^d$ were supposed up to now known but in practice also have to be estimated. In fact, they play an overriding role in the bias-variance ``tuning'' effect which motivated this work in Section~\ref{sec:motivation}.
%Looping over all candidates $m_j$ supposed inferior to a maximum number of levels $m_{\text{max}}$ (\textit{e.g.}\ to guarantee sparsity) would require to further expand the already intractable search space $\Q$ in~\eqref{eq:BICq} to $\Q \times \{1,\cdots,m_{\text{max}}\}^d$. 
By relying on the \textit{maximum a posteriori} procedure developed in Equation~(\ref{eq:ht}) \bleu{parallel to the neural network candidate generator}, we might drop a lot of unseen factor levels, \textit{e.g.}\ if $\s_{\ag_{j,h}}(x_{i,j}) \ll 1$ for all training observations $x_{i,j}$, the level $h$ ``vanishes'', \textit{i.e.}\ $\hat{\s}_{j,h} = 0$. Thus, it is not necessary to go through such a loop and in practice, we recommend to start with a user-chosen $\bm{m}=\boldsymbol{m}_{\max}$ and we will see in the experiments of Section~\ref{sec:experiments} that the proposed approach is able to explore small values of $\boldsymbol{m}$ and to select a value $\hat{\boldsymbol{m}}$ drastically smaller than $\boldsymbol{m}_{\max}$. This phenomenon, which reduces the computational burden of the quantization task, is also illustrated in the next section. \bleu{The hyper-parameter $\boldsymbol{m}_{\max}$ is problem-dependent and should be adjusted by the practitioner to meet his/her interpretability requirements.}

\section{Numerical experiments} \label{sec:experiments}

This section is divided into three complementary parts to assess the validity of our proposal, that we call hereafter \textit{glmdisc}. First, simulated data are used to evaluate its ability to recover the true data generating mechanism. Second, the predictive quality of the new learned representation approach is illustrated on several classical benchmark datasets from the UCI library. Third, we use it on \textit{Credit Scoring} datasets provided by Cr\'edit Agricole Consumer Finance (CACF), a major European company in the consumer credit market. The Python notebooks of all experiments, excluding the confidential real data of CACF, \bleu{are available online\footnote{\url{https://adimajo.github.io}}}.

\subsection{Simulated data: empirical consistency and robustness}

We focus here on discretization of continuous features (similar experiments could be conducted on categorical ones). Two continuous features $x_1$ and $x_2$ are sampled from the uniform distribution on $[0,1]$ and discretized using
\[\q_1(\cdot)=\q_2(\cdot) = (\mathds{1}_{(-\infty,1/3]}(\cdot),\mathds{1}_{(1/3,2/3]}(\cdot),\mathds{1}_{(2/3,\infty]}(\cdot)).\]
Here, following (\ref{eq:Cjhcont}), we have $d=2$ and $m_1=m_2=3$ and the cutpoints are $c_{j,1}=1/3$ and $c_{j,2}=2/3$ for $j=1,2$. Setting $\bth=(0,-2,2,0,-2,2,0)$, the target feature $y$ is then sampled from $p_{\bth}(\cdot | \q(\bm{x}))$ via the logistic model (\ref{eq:reglogq}).

\begin{figure*}
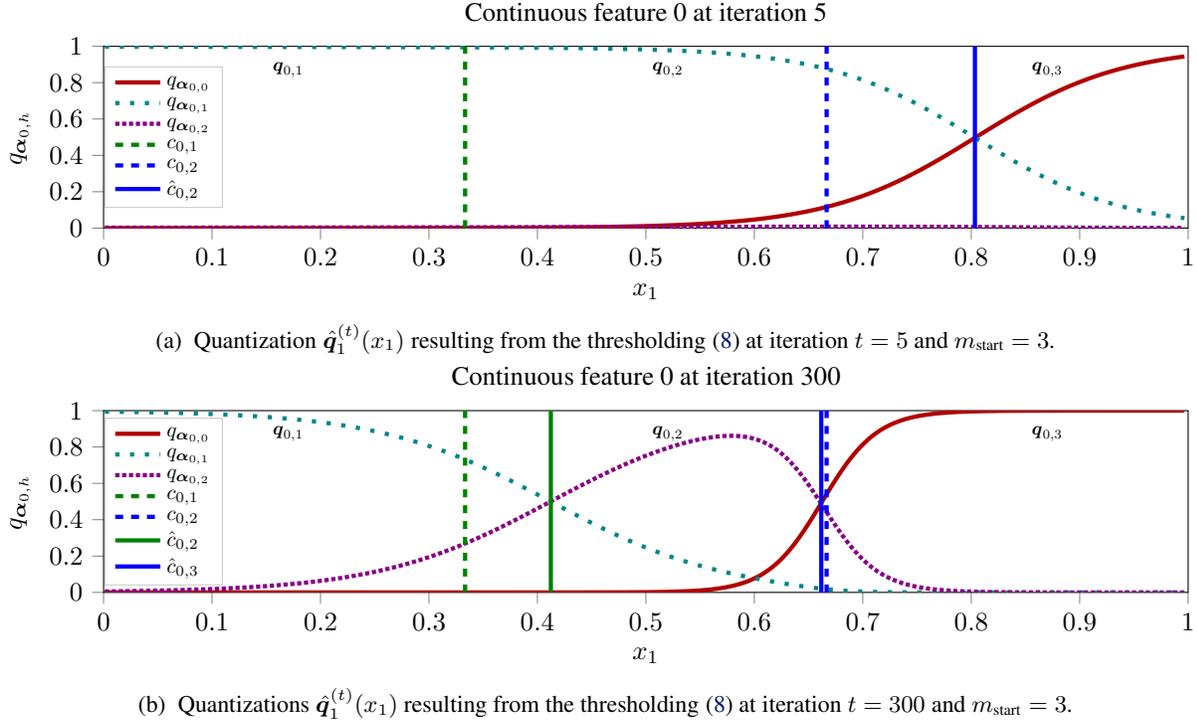

    \centering
    \begin{subfigure}[t]{\textwidth}
        \centering
        \input{True_simulated_data/feature_0_iteration_5.tex}
        \caption{\label{fig:MAPa} Quantization $\hat{\q}^{(t)}_1(x_1)$ resulting from the thresholding (\ref{eq:ht}) at iteration $t = 5$ and $m_{\text{start}} = 3$.}
    \end{subfigure}%
    
    \begin{subfigure}[t]{\textwidth}
        \centering
        \input{True_simulated_data/feature_0_iteration_300.tex}
        \caption{\label{fig:MAPb} Quantizations $\hat{\q}^{(t)}_1(x_1)$ resulting from the thresholding (\ref{eq:ht}) at iteration $t = 300$ and $m_{\text{start}} = 3$.}
    \end{subfigure}
    
    \caption{\label{fig:MAP} Quantizations $\hat{\q}^{(t)}_1(x_1)$ of experiment (A) resulting from the thresholding (\ref{eq:ht}).}
\end{figure*}

From the \textit{glmdisc} algorithm, we studied three cases:
\begin{enumerate}[(A)]
    \item First, the quality of the cutoff estimator $\hat{c}_{j,2}$ of $c_{j,2} = 2/3$ is assessed when the starting maximum number of intervals per discretized continuous feature is set to its true value $m_1=m_2= 3$;
    \item Second, we estimated the number of intervals $\hat{m}_1$ of $m_1=3$ when the starting maximum number of intervals per discretized continuous feature is set to $m_{\text{max}} = 10$; 
    \item Last, we added a third feature $x_3$ also drawn uniformly on $[0,1]$ but uncorrelated to $y$ and estimated the number $\hat{m}_3$ of discretization intervals selected for $x_3$. The reason is that a non-predictive feature which is discretized or grouped into a single value is \textit{de facto} excluded from the model, and this is a positive side effect.
\end{enumerate}
From a statistical point of view, experiment (A) assesses the empirical consistency of the estimation of $C_{j,h}$ motivated in Section~\ref{subsubsec:stochastic}, whereas experiments (B) and (C) focus on the consistency of the estimation of $m_j$ motivated in Section~\ref{subsubsec:choosing}. The results are summarized in Table~\ref{tab:estim_precision} where either 95\% confidence intervals \bleu{(\cite{sun2014fast}, hereafter CI)} or bar plots are given, with a varying sample size. Two iterations of experiment (A) are displayed on Figure~\ref{fig:MAP}: at first (Figure~\ref{fig:MAPa}), the proposed neural network fails to recover the true underlying discretization but after 300 iterations (Figure~\ref{fig:MAPb}), the ``smooth'' discretization $\q_{\ag}$ and its \textit{maximum a posteriori} $\hat{\q}$ get closer to the data generating mechanism, resulting in a very good estimation of $c_{j,2}$ (Column (A) of Table~\ref{tab:estim_precision}). Also note that the slight underestimation ($\hat{m}_1 = 2$ for 9 experiments out of 100) in (B) for $n=1{,}000$ is a classical consequence of the BIC criterion on small samples. As for (C) and as expected, spurious correlations with a small sample allow $x_3$ to enter the model with either $\hat{m}_3=2$ intervals (32 experiments out of 100) or $\hat{m}_3=3$ intervals (8 experiments out of 100). However, with a larger sample, feature $x_3$ is rightfully omitted from the final model, \textit{i.e.}\ with $\hat{m}_3=1$ interval (88 experiments out of 100).
\begin{table}[ht]
    \centering
    \caption{For different sample sizes $n$, (A) CI of $\hat{c}_{j,2}$ for $c_{j,2} = 2/3$. (B) Bar plot of $\hat{m} = 2, 3, 4$ (resp.) for $m_1=3$. (C) Bar plot of $\hat{m}_3 = 1, 2, 3$ (resp.) for $m_3=1$.}
    \label{tab:estim_precision}
\begin{tabular}{llllll}
$n$ & (A) $\hat{c}_{j,2}$ & (B) & $\hat{m}_1$ & (C) & $\hat{m}_3$ \\
\hline
$1{,}000$ & $[0.656,0.666]$ & \myobar{9}{90}{1} & \mybar{60}{32}{8} \\
$10{,}000$ & $[0.666,0.666]$ & \myobar{0}{100}{0} & \mybar{88}{12}{0}
\end{tabular}
\end{table}

 \newlength\figureheight
 \newlength\figurewidth
 \setlength\figureheight{4cm}
 \setlength\figurewidth{16cm}
 
\subsection{Benchmark data}

To test further the effectiveness of \textit{glmdisc} in a predictive setting, we gathered 6 datasets from the UCI library: the Adult dataset ($n=48{,}842$, $d=14$), the Australian dataset ($n=690$, $d=14$), the Bands dataset ($n=512$, $d=39$), the Credit-screening dataset ($n=690$, $d=15$), the German dataset ($n=1{,}000$, $d=20$) and the Heart dataset ($n=270$, $d=13$). Each of these datasets has mixed (continuous and categorical) features and a binary response to predict. To get more information about these datasets, their respective features, and the predictive task associated with them, the interested reader may refer to the UCI website\footnote{\cite{Dua:2017} : http://archive.ics.uci.edu/ml}.

Now that we made sure that our approach is empirically consistent, \textit{i.e.}\ it is able to find the true quantization in a well-specified setting, we wish to verify now that embedding the learning of a good quantization in the predictive task \textit{via glmdisc} is better than other methods that rely on \textit{ad hoc} criteria. As we were primarily interested in logistic regression, we will compare our approach to a ``na\"{\i}ve'' additive linear logistic regression (\bleu{on non-quantized features} - hereafter ALLR), a logistic regression on continuous discretized data using the now standard MDLP algorithm from~\cite{fayyad1993multi} and categorical grouped data using $\chi^2$ tests of independence between each pair of factor levels and the target in the same fashion as the ChiMerge discretization algorithm proposed by~Kerber~\yrcite{kerber1992chimerge} (hereafter MDLP/$\chi^2$). As the original use case stems from \textit{Credit Scoring}, we use the performance metric usually monitored by \textit{Credit Scoring} practitioners, which is the Gini coefficient, directly related to the Area Under the ROC Curve (Gini $= 2 \times \text{AUC} -1$). \bleu{In this Section and the next, Gini indices are reported on a random 30~\% test set.} Table~\ref{tab:banchmark} shows our approach yields significantly better results on these rather small datasets where the added flexibility of quantization might help the predictive task.

\begin{table}
    \centering
        \caption{Gini indices (the greater the value, the better the performance) of our proposed quantization algorithm \textit{glmdisc} and two baselines: ALLR and MDLP / $\chi^2$ tests obtained on several benchmark datasets from the UCI library.}
    \label{tab:banchmark}
\begin{tabular}{llll}
Dataset & ALLR & MDLP/$\chi^2$ & \textit{glmdisc} \\
\hline
Adult & 81.4 (1.0) & \bf{85.3} (0.9) & 80.4 (1.0) \\
Australian & 72.1 (10.4) & 84.1 (7.5) & \bf{92.5} (4.5) \\
Bands & 48.3 (17.8) & 47.3 (17.6) & \bf{58.5} (12.0) \\
Credit & 81.3 (9.6) & 88.7 (6.4) & \bf{92.0} (4.7) \\
German & 52.0 (11.3) & 54.6 (11.2) & \bf{69.2} (9.1) \\
Heart & 80.3 (12.1) & 78.7 (13.1) & \bf{86.3} (10.6)
\end{tabular}
\end{table}

\subsection{\textit{Credit Scoring} data}

Discretization and grouping are preprocessing steps relatively ``manually'' performed in the field of \textit{Credit Scoring}, using $\chi^2$ tests for each feature or so-called Weights of Evidence~\cite{zeng2014necessary}. This back and forth process takes a lot of time and effort and provides no particular statistical guarantee.

Table~\ref{tab:real_data} shows Gini coefficients of several portfolios for which there are $n=50{,}000$, $n=30{,}000$, $n=50{,}000$, $n=100{,}000$, $n=235{,}000$ and $n=7{,}500$ clients respectively and $d=25$, $d=16$, $d=15$, $d=14$, $d=14$ and $d=16$ features respectively. Approximately half of these features were categorical, with a number of factor levels ranging from $2$ to $100$. 

We compare the rather manual, in-house approach that yields the current performance, the na\"{\i}ve additive linear logistic regression (ALLR) and \textit{ad hoc} methods (MDLP/$\chi^2$) introduced in the previous section to our \textit{glmdisc} proposal. Beside the classification performance, interpretability is maintained and unsurprisingly, the learned representation comes often close to the ``manual'' approach: for example, the complicated in-house coding of job types is roughly grouped by \textit{glmdisc} into \textit{e.g.}\ ``worker'', ``technician'', \textit{etc.}
Our approach shows approximately similar results than MDLP/$\chi^2$, potentially due to the fact that contrary to the two previous experiments with simulated or UCI data, the classes are imbalanced ($< 3 \%$ defaulting loans), which would require special treatment while back-propagating the gradients~\cite{anand1993improved}. Note however that it is never significantly worse; for the Electronics dataset and as was the case for most UCI datasets, \textit{glmdisc} is significantly superior, which in the \textit{Credit Scoring} business might end up saving millions to the financial institution.

\begin{table}
    \centering
        \caption{Gini indices (the greater the value, the better the performance) of our proposed quantization algorithm \textit{glmdisc}, the two baselines of Table~\ref{tab:banchmark} and the current scorecard (manual / expert representation) obtained on several portfolios of Cr\'edit Agricole Consumer Finance.}
    \label{tab:real_data}
%\begin{tabular}{lp{0.191\linewidth}p{0.138\linewidth}p{0.13\linewidth}p{0.15\linewidth}}
\resizebox{.5\textwidth}{!}{\begin{tabular}{lllll}
Portfolio & ALLR & Current & MDLP/$\chi^2$ & \textit{glmdisc} \\
\hline
Automobile & \bf{59.3} (3.1) & 55.6 (3.4) & \bf{59.3} (3.0) & 59.1 (3.0) \\
Renovation & 52.3 (5.5) & 50.9 (5.6) & 54.0 (5.1) & \bf{56.7} (4.8) \\
Standard & 39.7 (3.3) & 37.1 (3.8) & \bf{45.3} (3.1) & 44.0 (3.1) \\
Revolving & 62.7 (2.8) & 58.5 (3.2) & \bf{63.2} (2.8) & 62.3 (2.8) \\
Mass retail & 52.8 (5.3) & 48.7 (6.0) & 61.4 (4.7) & \bf{61.8} (4.6) \\
Electronics & 52.9 (11.9) & 55.8 (10.8) & 56.3 (10.2)  & \bf{72.6} (7.4)
\end{tabular}}
\end{table}

%\subsection{Complexity and running time}

Regarding complexity, there are at most $\mathcal{O}(m_j^2)$ $\chi^2$ tests performed in all benchmarks for categorical features as initially, all pairwise tests have to be computed. The MDLP algorithm has to first sort the training samples ($\mathcal{O}(n \ln{n})$ operations) and then recursively assess the entropy produced by cutting at each ``boundary point'', \textit{i.e.} where consecutive training points, say $x_{i,j},x_{i',j}$, have different targets ($y_i \neq y_{i'}$). There are $\mathcal{O}(b_j^2)$ such operations where $b_j$ is the number of these ``boundary points''~\cite{ramirez2016data}. Our approach, the \textit{glmdisc} algorithm, requires that we fit a softmax with $m_j$ output classes per feature and training epoch $(t)$ \bleu{which is quite low.  About the length of the gradient ascent chain, there is no stopping rule except the time budget T. However, the required T value to obtain relevant candidates is low: approx.\ 20-40 iterations for the experiments of Section~\ref{sec:experiments}. Figure~\ref{fig:MAP} uses a small learning rate to showcase both the empirical consistency of the relax and the effect of the MAP scheme in exploring a lower number of quantization levels $m_j$.}
On Google Collaboratory, and relying on Keras~\cite{chollet2015keras} and Tensorflow~\cite{tensorflow2015-whitepaper} as a backend, it took less than an hour to perform discretization and grouping for each dataset of Table~\ref{tab:real_data}, making it in this regard also comparable to MDLP/$\chi^2$ methods.

\section{Concluding remarks} \label{sec:conclusion}

Feature quantization (discretization for continuous features, grouping of factor levels for categorical ones) in a supervised multivariate classification setting is a recurring problem in many industrial contexts. It was formalized as a highly combinatorial representation learning problem and a new algorithmic approach, named \textit{glmdisc}, has been proposed as a sensible approximation of a classical statistical information criterion.

This algorithm relies on the use of a softmax approximation of each discretized or grouped feature. This proposal can alternatively be replaced by any other univariate multiclass predictive model, which makes it flexible and adaptable to other problems. Prediction of the target feature, given quantized features, was exemplified with logistic regression, although here as well, it can be swapped with any other supervised classification model, provided it is the same as the output layer of the proposed neural network. \bleu{Thus, the extension to penalized logistic regression or any Generalized Linear Model is straightforward.} Its good computational properties were put to use while maintaining the interpretability necessary to some fields of application.

The experiments showed that, as was sensed empirically by statisticians in the field of \textit{Credit Scoring}, discretization and grouping can indeed provide better models than standard logistic regression. This novel approach allows practitioners to have a fully automated and statistically well-grounded tool that achieves better performance than \textit{ad hoc} industrial practices at the price of decent computing time but much less of the practitioner's valuable time. As a rule of thumb, a month is generally allocated to data pre-processing for a single data scientist working on a single scorecard that can now be invested in tasks that add more value, \textit{e.g.}\ more data, better data quality.

As described in the introduction, logistic regression is additive in its inputs which does not allow to take into account conditional dependency, as stated by Berry et al.\ \yrcite{berry2010testing}. This problem is often dealt with by sparsely introducing ``interactions'', \textit{i.e.}\ products of two (pairwise interactions) or more features. This leads again to a model selection challenge on a highly combinatorial discrete space that could be solved with a similar approach. In a broader context with no restriction on the predictive model, Tsang et al.\ \yrcite{tsang2018detecting} already made use of neural networks to estimate the presence or absence of statistical interactions. The parsimonious addition of pairwise interactions among quantized features, that might influence the quantization process introduced in this work, is a future area of research.

\bibliography{feature_quantization}
\bibliographystyle{icml2019}

\end{document}